\documentclass[article]{eptcs}

\providecommand{\event}{LSFA 2011}

\usepackage{breakurl}
\usepackage{amsfonts}
\usepackage{amsthm}
\usepackage{amsmath}
\usepackage{amssymb}

\newtheorem{te}{Theorem}[section]
\newtheorem{la}{Lemma}[section]
\newtheorem{co}{Corolary}[section]
\newtheorem{po}{Proposition}[section]
\newtheorem{de}{Definition}[section]
\newtheorem{ex}{Example}[section]
\newtheorem{re}{Remark}[section]
\newtheorem{no}{Notation}[section]

\newcommand{\bpr}{\begin{proof}}
\newcommand{\epr}{\end{proof}}
\newcommand{\bte}{\begin{te}}
\newcommand{\ete}{\end{te}}
\newcommand{\bla}{\begin{la}}
\newcommand{\ela}{\end{la}}
\newcommand{\bco}{\begin{co}}
\newcommand{\eco}{\end{co}}
\newcommand{\bpo}{\begin{po}}
\newcommand{\epo}{\end{po}}
\newcommand{\bde}{\begin{de}}
\newcommand{\ede}{\end{de}}
\newcommand{\bex}{\begin{ex}}
\newcommand{\eex}{\end{ex}}
\newcommand{\bre}{\begin{re}}
\newcommand{\ere}{\end{re}}
\newcommand{\bno}{\begin{no}}
\newcommand{\eno}{\end{no}}

\title{Classical and quantum satisfiability\thanks{This research was
    supported by Fapesp Thematic Projects 2008/03995-5 (LogProb) and
    2010/51038-0(LogCons).}}  \author{Anderson de Ara\'ujo\thanks{Partially supported by FAPESP grant 2011/07781-2.} 
  \institute{Institute of Mathematics and Statistics\\
    University of S\~ao Paulo\\
    S\~ao Paulo, Brazil} \email{aaraujo@ime.usp.br} \and \qquad Marcelo
  Finger\thanks{Partially supported by CNPq grant PQ 302553/2010-0.}
  \institute{Institute of Mathematics and Statistics\\
    University of S\~ao Paulo\\
    S\~ao Paulo, Brazil} \email{\quad mfinger@ime.usp.br} }

\def\titlerunning{Classical and quantum satisfiability}

\def\authorrunning{A. de Ara\'ujo \& M. Finger}

\begin{document}

\maketitle

\sloppy

\begin{abstract}
We present the linear algebraic definition of QSAT and propose a direct logical characterization of such a definition. 
We then prove that this logical version of QSAT is not an extension of classical satisfiability problem (SAT). This shows that 
QSAT does not allow a direct comparison between the complexity classes NP and QMA, for which SAT and QSAT are respectively complete.
\end{abstract}

\section{Introduction}

Quantum computation is the paradigm of computer science wherein computations are treated as quantum physical processes. 
Basically, the interest on this paradigm relies on the possibility that some problems may be solved more efficiently by quantum 
computers than by classical ones (Cf. \cite{bennet1997a}). To analyze the relationship between the capabilities of these two very 
different kinds of computers, quantum versions of the classical computational complexity classes have been defined. In particular, 
the time-complexity classes $BQP$ and $QMA$ have received considerable attention (Cf. \cite[p. 201-234]{arora2009}).

Since quantum mechanics predicts probabilities of events (Cf. \cite{cohen-tannoudji1977}), $BQP$ and $QMA$ are generalizations of 
probabilistic classes. $BQP$ is the class of problems decidable in polynomial time with bounded error on a quantum computer; it is 
the quantum generalization of $BPP$, which is, in turn, the probabilistic version of $P$. $QMA$ is the quantum-Merlin-Arthur complexity class, 
the class of decision problems that can be efficiently verified by a quantum computer; it is the quantum version of the class $MA$, which is the 
classical probabilistic generalization of $NP$.

$NP$-completeness is an important phenomena in the understanding of the limits between the classes $P$ and $NP$.  In the case of $BQP$ and $QMA$, 
the same can be said about $QMA$-completeness. The first $QMA$-complete problem was formulated by Kitaev and it is called \emph{local 
Halmiltonian satisfiability problem} ($HSAT$); it can be found in \cite[p. 142]{kitaev2002}. $HSAT$ is a generalization of the \emph{MAX-SAT}
 problem to context of quantum mechanics, where Hamiltonian matrices have a central role in the description of physical systems. In \cite{bravyi2006}, 
Bravyi changed some aspects of $HSAT$ in order to obtain a quantum version of the $SAT$ problem. Bravy's version of $HSAT$ is called $QSAT$ and, 
in order to make explicit its logical core, in \cite{bravyi2010} Bravyi et al. define $QSAT$ in the following way:

\begin{description}

\item[Input:] A set of reduced density matrices $\{((I_{k} - | v \rangle \langle v |) \otimes I_{n-k})_{1}, \ldots, ((I_{k} - | v 
\rangle \langle v |) \otimes I_{n-k})_{m}\}$ on the Hilbert space of $n$ qubits is given, where $| v \rangle$ is a vector in the 
$2^{k}$-dimensional Hilbert space of some $k$-tuple of qubits, $I_{k}$ is the identity on that Hilbert space, and $I_{n-k}$ is the identity 
on the remaining qubits.

\item[Problem:] Is there a vector $|w\rangle$ in the Hilbert space of $n$ qubits such that

\begin{center}
$\langle w |((I_{k} - | v \rangle \langle v |) \otimes I_{n-k})_{i}| w \rangle = 0$ for all $i$ with $1 \leq i \leq m$?
\end{center}

Or, for all $|w\rangle$ in the Hilbert space of $n$ qubits,

\begin{center}
$\langle w |((I_{k} - | v \rangle \langle v |) \otimes I_{n-k})_{i}| w \rangle \geq \epsilon$ for some $i$ with $1 \leq i \leq m$, 
where $\epsilon = \Omega (1 / poly(n))$ is a fixed real number\footnote{It is necessary to fix such an $\epsilon$ in order to exclude 
the cases in which there exists no exact satisfying vector, but there are approximate $| w \rangle$ such that equations 
$\langle w |((I_{k} - | v \rangle \langle v |) \otimes I_{n-k})_{i}| w \rangle = 0$ are satisfied with an exponentially small error.}?
\end{center}

\end{description}

The idea underlining the formulation of $QSAT$ is that, given a propositional sentence $\phi = \psi_{1} \wedge \cdots \wedge \psi_{m}$ in 
conjunctive normal form, the vector $|v\rangle$ in each reduced density matrix $((I_{k} - | v \rangle \langle v |) \otimes I_{n-k})_{j}$ in a 
$QSAT$ problem corresponds to a classical evaluation $v$ that satisfies all clauses $\psi_{j}$ of $\phi$. Given that $I_{k} - | v \rangle 
\langle v |$ is part of each $((I_{k} - | v \rangle \langle v |) \otimes I_{n-k})_{j}$, if there is a vector $|w\rangle$ as above, $|w\rangle$ 
is orthogonal to each of these reduced density matrices and so $|w\rangle$ corresponds to an evaluation $w$ that satisfies $\phi$.

Clearly this is a quantum view about $SAT$. Moreover, Bravyi showed in \cite{bravyi2006} that $QSAT$ is $QMA$-complete when the number of qubits 
$n$ is greater then 2. For this reason $QSAT$ has drawn attention in the literature about quantum computational complexity (Cf. \cite{laumann2010}): 
it is a $QMA$-complete problem that is related to an $NP$-complete problem. However, the relationship between complexity classes $NP$ and $QMA$ is not 
very well understood. This relationship apparently involves more than mere extensions of problems with probabilities. The probabilistic satisfiability 
problem (PSAT) is a problem that clearly
extends SAT, but it was shown to remain NP-complete problem \cite{georgakopoulos1988}. In \cite{liu2006}, a variation of $QSAT$ more closely related 
to $PSAT$ than to $SAT$ was presented. In
\cite{laumann2010}, stochastic versions of $QSAT$ was explored. But no relationship between instances of SAT, PSAT and of QSAT was established.

The present paper will show that the idea underlining $QSAT$, and which permits us to think it as a generalization of $SAT$, is not adequate, from a 
logical perspective, to compare the classes $NP$ and $QMA$. More precisely, the aim of this paper is to show that, when $QSAT$ is formalized in order 
to establish connections with $SAT$, there are evaluations that satisfies $SAT$ but which do not directly correspond to matrices in the form that $QSAT$ 
is defined. In Section 2, $QSAT$ will be formulated from $SAT$, using the notion of \emph{quantum assignment}. Given this, in Section 3, it will be 
proved that $QSAT$ in terms of quantum assignments does not correspond to $SAT$, that is to say, $SAT$ cannot be viewed as a subcase of $QSAT$. Since 
quantum assignments are a very natural way of defined $QSAT$ from $SAT$, the main result of this paper shows that $QSAT$ is not a good problem to 
analyze the relationship between $NP$ and $QMA$.

\section{Classical and quantum satisfiability}

In this section, from the definition of $SAT$ it will be provided a logical version of $QSAT$. For this end, let $X$ be a set of 
\emph{propositional variables}. Consider $L$ the \emph{propositional language} defined over $X$ using the \emph{alphabet} 
$\{\neg, \vee, \wedge\}$. An $L$-formula $\phi$ is in \emph{conjunctive normal form} (CNF) if $\phi = \psi_{1} \wedge \cdots \wedge \psi_{m}$ 
and, for each $i$, $$\psi_{i} = \chi_{1} \vee \cdots \vee \chi_{k},$$ where $\chi_{j} \in \{x, \neg x\}$ for $x \in X$. Besides this, 
if the propositional variables of $\phi$ are in the set $var(\phi) = \{x_{1},\ldots,x_{n}\}$, $\phi$ is called an $L$-formula with 
\emph{dimension} $(k,n)$.

\bde\label{sat}
An \emph{evaluation assignment} is a function $v$ from $X$ to $\{0,1\}$. An evaluation assignment $v$ is extended to a \emph{full} 
evaluation assignment $\hat{v}: L \to \{0,1\}$ in the usual way: $\hat{v}(x) = v(x)$ for $x \in X$; $\hat{v}(\neg \phi) = 1$ if, 
and only if, $\hat{v}(\phi) = 0$; $\hat{v}(\phi \vee \psi) = 1$ if, and only if, $\hat{v}(\phi) = 1$ or $\hat{v}(\psi) = 1$; 
$\hat{v}(\phi \wedge \psi) = 1$ if, and only if, $\hat{v}(\phi) = 1$ and $\hat{v}(\psi) = 1$. An $L$-formula $\phi$ is \emph{satisfiable} 
when there is an evaluation assignment $v$ such that $\hat{v}(\phi) = 1$. The \emph{$k$-satisfatibility problem} ($k$-$SAT$) is the following 
question: \emph{Given an $L$-formula $\phi$ in CNF with dimension $(k,n)$, is $\phi$ satisfiable?}
\ede

In the definition of $SAT$, the meaning of an $L$-formula $\phi$ was defined in terms of evaluation assignments. In order to provide a 
quantum interpretation of the meaning of $\phi$, a natural way to proceed is to convert evaluation assignments into density matrices, because 
in the density operator formulation of quantum mechanics there is a postulate that establishes which to each body in an isolated physical 
systems corresponds a density operator in a Hilbert space \cite{cohen-tannoudji1977}. The formulation of $QSAT$ exhibited in the Introduction 
relies on this intuition; in what follows such a perspective will be situated in a logical context.

Given an $L$-formula $\phi$ such that $var(\phi) = \{x_{1},\ldots,x_{n}\}$, the Hilbert space associate to $\phi$ is the vector space
 $\mathbb{C}^{\otimes n}_{2}$ of dimension $2^{n}$ defined on the complex field $\mathbb{C}$ such as in \cite[p. 61]{nielsen2000}. 
The \emph{computational base} of $\mathbb{C}^{\otimes n}_{2}$ is the basis set $\{|b^{1}\rangle, \ldots, |b^{2^{n}}\rangle\}$ where 
each vector is defined as $$|b^{k}\rangle = \begin{pmatrix} b^{k}_{1} \\ \vdots \\ b^{k}_{2^{n}} \end{pmatrix},$$

for $b^{k}_{i} = \left\{
\begin{array}{rcl}
1 & \mbox{if $k =i$},\\
0 & \mbox{otherwise}.
\end{array}\right.$

 The Hilbert space $\mathbb{C}^{\otimes n}_{2}$ has an inner product $\langle \: | \: \rangle : \mathbb{C}^{\otimes n}_{2} 
\times \mathbb{C}^{\otimes n}_{2} \to \mathbb{C}^{\otimes n}_{2}$ defined in the following way:

$$\langle v | w \rangle = \begin{pmatrix} v^{*}_{1} & \cdots & v^{*}_{2^{n}} \end{pmatrix} \begin{pmatrix} w_{1} \\ \vdots \\ w_{2^{n}} \end{pmatrix},$$

where $v^{*}_{i}$ is the complex conjugate of $v_{i}$. From this, it is possible to define a logical version of $QSAT$ into $\mathbb{C}^{\otimes n}_{2}$ 
in accordance with the definition given in \cite{bravyi2010}.

\bde\label{qsat}
For each clause $\psi_{i}$ of an $L$-formula $\phi = \psi_{1} \wedge \cdots \wedge \psi_{m}$ in CNF such that $\psi_{i} = \chi_{1} \vee \cdots \vee 
\chi_{k}$ and $var(\phi) = \{x_{1},\ldots,x_{n}\}$, a \emph{quantum assignment to $\psi_{i}$} is a $2^{n} \times 2^{n}$-matrix $\psi^{(\hat{v})}_{i}$ 
such that $$\psi^{(\hat{v})}_{i} = a(I_{k} - | v(x_{i_{1}}) \cdots v(x_{i_{k}}) \rangle \langle v(x_{i_{1}}) \cdots v(x_{i_{k}}) |) \otimes I_{n-k},$$

where $a$ is some polynomial-time computable complex number in $\mathbb{C}$, $v \in Eval(\phi)$ is such that, for all $i$ with $0 \leq i \leq m$, 
$\hat{v}(\psi_{i}) = 1$, and $x_{i_{1}}, \ldots, x_{i_{k}}$ are the propositional variables in $var(\phi)$ that occur in $\psi_{i}$. Fix a real 
number $\epsilon = \Omega (1 / poly(n))$. Thus, $\phi$ is \emph{quantum satisfiable} if there is a vector $|w\rangle$ in $\mathbb{C}^{\otimes n}_{2}$ 
such that

\begin{center}
$\langle w | \psi^{(\hat{v})}_{i} | w \rangle = 0$ for all $i$ with $1 \leq i \leq m$;
\end{center}

otherwise, $\phi$ is \emph{quantum unsatisfiable}, i.e., for all $|w\rangle$ in $\mathbb{C}^{\otimes n}_{2}$,

\begin{center}
$\langle w | \psi^{(\hat{v})}_{i} | w \rangle \geq \epsilon$ for some $i$ with $1 \leq i \leq m$.
\end{center}

The \emph{quantum $k$-satisfatibility problem} ($k$-$QSAT^{l}$) is the following question: \emph{Given an $L$-formula $\phi$ in CNF with 
dimension $(k,n)$, is $\phi$ quantum satisfiable?}
\ede

It is important to note that $QSAT^{l}$ is a restriction of the original problem $QSAT$ shown in the Introduction. As explained above, the 
relationship between $QSAT$ and $SAT$ is established at an informal and intuitive level, but in $QSAT^{l}$ the reduced density matrices are quantum 
assignments, which are matrices constructed from evaluation assignments. In other words, $QSAT^{l}$ is a logical version of $QSAT$ defined directly 
from $SAT$. Hence, it is possible now to evaluate the relationship between $QSAT$ and $SAT$ from a logical point of view, looking at the relationship between $QSAT^{l}$ and $SAT$.

\section{$k$-$SAT$ and $k$-$QSAT^{l}$}

In this section it will be shown that, although all problems in $QSAT^{l}$ are just quantum versions of problems in $SAT$, the conversion of a 
solution to a problem in $SAT$ not necessarily corresponds to a solution of the same problem in $QSAT^{l}$. Since $QSAT^{l}$ is a logical 
restriction of $QSAT$, this means that $QSAT$ could be considered a quantum generalization of $SAT$ at an intuitive level, but from a logical 
perspective the relationship between $QSAT$ and $SAT$ is week. Indeed, given definitions \ref{sat} and \ref{qsat}, it seems reasonable to consider 
$QSAT$ a good generalization of $SAT$ only if each solution to an instance of a $k$-$SAT$ problem can be translated into a solution to an instance 
of a $k$-$QSAT^{l}$ problem, this section shows that this is not the case.

More precisely, let $\phi$ be an $L$-formula in CNF with dimension $(k,n)$. To provide a positive solution to the $k$-$SAT$ problem relative to $\phi$ 
means to find an evaluation $v$ such that $\hat{v}(\phi) = 1$. If $QSAT$ is a good generalization of $SAT$, then, for each $v$ such that 
$\hat{v}(\phi) = 1$, it should be possible to find a vector $|w\rangle$ in $\mathbb{C}^{\otimes n}_{2}$ such that, first, 
$\langle w | \psi^{(\hat{v})}_{i} | w \rangle = 0$ for all $i$ with $1 \leq i \leq m$ and, second, $v$ can be directly translated into 
$|w\rangle$. Certainly, supposing that $var(\phi) = \{x_{1}, \ldots, x_{n}\}$, a very \emph{natural conversion} of such an evaluation $v$ 
is just the vector $| v(x_{1}) \cdots v(x_{n})\rangle$, i.e., $|w\rangle = | v(x_{1}) \cdots v(x_{n})\rangle$ should be a vector that is orthogonal 
to the quantum assignments associated to the clauses of $\phi$ because $\hat{v}$ satisfies $\phi$. Nevertheless, consider the following example.

\bex\label{example}
Take the $L$-formula $\phi = (x \vee \neg y) \wedge (\neg x \vee z)$. The evaluation $v \in Eval(\phi)$ such that $v(x) = 1$, $v(y) = 0$ and $v(z) = 1$ 
is such that $\hat{v}(x \vee \neg y) = \hat{v}(\neg x \vee z) = 1$ and so $\hat{v}(\phi) = 1$. In this case,
 $$| v(x) v(y) \rangle \langle v(x) v(y) | = || 1 \rangle \otimes | 0 \rangle \rangle \langle | 1 \rangle \otimes | 0 \rangle| = 
\begin{pmatrix} 0 & 0 & 0 & 0 \\ 0 & 0 & 0 & 0 \\ 0 & 0 & 1 & 0 \\ 0 & 0 & 0 & 0 \end{pmatrix}$$ 
\begin{center}
and
\end{center}
 $$| v(x) v(z) \rangle \langle v(x) v(z) | = || 1 \rangle \otimes | 1 \rangle \rangle \langle | 1 \rangle \otimes | 1 \rangle| = 
\begin{pmatrix} 0 & 0 & 0 & 0 \\ 0 & 0 & 0 & 0 \\ 0 & 0 & 0 & 0 \\ 0 & 0 & 0 & 1 \end{pmatrix}.$$

As $k = 2$ and $n = 3$, $(I_{k} - | v(x) v(y) \rangle \langle v(x) v(y) |) \otimes I_{n-k}$ and $(I_{k} - | v(x) v(z) \rangle \langle v(x) v(z) |)
 \otimes I_{n-k}$ are, respectively, the following matrices 

\begin{center}
$a\begin{pmatrix} 1 & 0 & 0 & 0 & 0 & 0 & 0 & 0 \\ 0 & 1 & 0 & 0 & 0 & 0 & 0 & 0 \\ 0 & 0 & 1 & 0 & 0 & 0 & 0 & 0 \\ 
0 & 0 & 0 & 1 & 0 & 0 & 0 & 0 \\ 0 & 0 & 0 & 0 & 0 & 0 & 0 & 0 \\ 0 & 0 & 0 & 0 & 0 & 0 & 0 & 0 \\ 0 & 0 & 0 & 0 & 0 & 0 & 1 & 0 \\ 
0 & 0 & 0 & 0 & 0 & 0 & 0 & 1 \end{pmatrix}$, $b\begin{pmatrix} 1 & 0 & 0 & 0 & 0 & 0 & 0 & 0 \\ 0 & 1 & 0 & 0 & 0 & 0 & 0 & 0 \\
 0 & 0 & 1 & 0 & 0 & 0 & 0 & 0 \\ 0 & 0 & 0 & 1 & 0 & 0 & 0 & 0 \\ 0 & 0 & 0 & 0 & 1 & 0 & 0 & 0 \\ 0 & 0 & 0 & 0 & 0 & 1 & 0 & 0 \\ 
0 & 0 & 0 & 0 & 0 & 0 & 0 & 0 \\ 0 & 0 & 0 & 0 & 0 & 0 & 0 & 0 \end{pmatrix}$.
\end{center}

However, $| v(x) v(y) v(z) \rangle = | 101 \rangle = | 1 \rangle \otimes | 0 \rangle \otimes | 1 \rangle$ is the  vector

$$\begin{pmatrix} 0 \\ 0 \\ 0 \\ 0 \\ 0 \\ 1 \\ 0 \\ 0 \end{pmatrix}.$$

Hence, $a(I_{k} - | v(x) v(y) \rangle \langle v(x) v(y) |) \otimes I_{n-k} | v(x) v(y) v(z) \rangle = 0$ but $b(I_{k} - | v(x) v(z) 
\rangle \langle v(x) v(z) |) \otimes I_{n-k} | v(x) v(y) v(z) \rangle \neq 0$ for any $a,b \in \mathbb{C}$.
\eex

This example shows that the natural conversion exhibited above does not work for a particular $L$-formula. The next result generalizes 
example \ref{example}.

\bpo\label{proposition}
Let $\phi$ be a satisfiable $L$-formula in CNF with dimension $(k,n)$ such that $var(\phi) = \{x_{1}, \ldots, x_{n}\}$. Suppose that $\psi_{p}$ 
and $\psi_{q}$ are clauses of $\phi$ such that $var(\psi_{p}) \neq var(\psi_{q})$. Then, there is an evaluation $v \in Eval(\phi)$ such that, 
for all $i$, $\hat{v}(\psi_{i}) = 1$ but
\begin{center}
either $\langle v(x_{1}) \cdots v(x_{n}) | \psi^{(\hat{v})}_{p} | v(x_{1}) \cdots v(x_{n}) \rangle \neq 0$ or $\langle v(x_{1}) \cdots v(x_{n}) |
 \psi^{(\hat{v})}_{q} | v(x_{1}) \cdots v(x_{n}) \rangle \neq 0$.
\end{center}
\epo
\bpr
Consider $\psi_{p} = \chi^{p}_{1} \vee \cdots \vee \chi^{p}_{k}$ and $\psi_{q} = \chi^{q}_{1} \vee \cdots \vee \chi^{q}_{k}$. Let $x^{p}_{1}, 
\ldots, x^{p}_{k}$ and $x^{q}_{1}, \ldots, x^{q}_{k}$ be the propositional variables in $var(\phi) = \{x_{1}, \ldots, x_{n}\}$ that occur, 
respectively, in $\psi_{p}$ and $\psi_{q}$. Note that $\langle v(x_{1}) \cdots v(x_{n}) | \psi^{(\hat{v})}_{i} | v(x_{1}) \cdots v(x_{n}) \rangle = 0$ 
if, and only if, $\psi^{(\hat{v})}_{i} | v(x_{1}) \cdots v(x_{n}) \rangle = ((I_{k} - | v(x^{i}_{1}) \cdots v(x^{i}_{k}) \rangle \langle
 v(x^{i}_{1}) \cdots v(x^{i}_{k}) |) \otimes I_{n-k}) | v(x_{1}) \cdots v(x_{n}) \rangle = 0$. Thus, it will be shown that there exists an 
evaluation $v \in Eval(\phi)$ such that, for all $i$, $\hat{v}(\psi_{i}) = 1$ but

\begin{center}
either $((I_{k} - | v(x^{p}_{1}) \cdots v(x^{p}_{k}) \rangle \langle v(x^{p}_{1}) \cdots v(x^{p}_{k}) |) \otimes I_{n-k}) | v(x^{p}_{1})
 \cdots v(x^{p}_{k}) \rangle \neq \vec{0}$ or $((I_{k} - | v(x^{q}_{1}) \cdots v(x^{q}_{k}) \rangle \langle v(x^{q}_{1}) \cdots v(x^{q}_{k}) |) 
\otimes I_{n-k}) | v(x^{q}_{1}) \cdots v(x^{q}_{k}) \rangle \neq \vec{0}$.
\end{center}

The matrix $(((I_{k} - | v(x^{p}_{1}) \cdots v(x^{p}_{k}) \rangle \langle v(x^{p}_{1}) \cdots v(x^{p}_{k}) |) \otimes I_{n-k})$ associated 
to $\psi_{p}$ is such that either $((I_{k} - | v(x^{p}_{1}) \cdots v(x^{p}_{k}) \rangle \langle v(x^{p}_{1}) \cdots v(x^{p}_{k}) |) 
\otimes I_{n-k}) | v(x^{p}_{1}) \cdots v(x^{p}_{k}) \rangle = 0$ or $((I_{k} - | v(x^{p}_{1}) \cdots v(x^{p}_{k}) \rangle \langle v(x^{p}_{1})
 \cdots v(x^{p}_{k}) |) \otimes I_{n-k}) | v(x^{p}_{1}) \cdots v(x^{p}_{k}) \rangle \neq 0$. Suppose that the first case is true. 
Since $var(\psi_{p}) \neq var(\psi_{q})$, without lost of generality, let $x^{q}_{j} \in \{x_{1}, \ldots, x_{n}\}$ be such that $x^{q}_{j} 
\in var(\psi_{q}) - var(\psi_{p})$. In this way, $x^{q}_{j} = \chi^{q}_{j}$ or $\neg x^{q}_{j} = \chi^{q}_{j}$ for some $j$ with $1 \leq j \leq k$, 
designate it just by $\chi^{q}_{j}$. Due to the hypothesis, $\phi$ is satisfiable, and so there is an evaluation $v \in Eval(\phi)$ such that
 $\hat{v}(\psi_{p}) = \hat{v}(\psi_{q}) = 1$. Take some $\chi^{p}_{i}$ for which $\hat{v}(\chi^{p}_{i}) = 1$ and consider the $x^{p}_{i}$ 
that occurs in $\chi^{p}_{i}$. Permute $\chi^{q}_{j}$ in $\psi_{q}$ until $\chi^{q}_{j}$ is the position $i$ in $\psi_{q}$, i.e., do the 
following: $\chi^{q}_{1} \vee \cdots \vee \chi^{q}_{j-1} \vee \chi^{q}_{j+1} \vee \chi^{q}_{j} \vee \cdots \vee \chi^{q}_{k}$, $\chi^{q}_{1} 
\vee \cdots \vee \chi^{q}_{j-1} \vee \chi^{q}_{j+1} \vee \chi^{q}_{j+2} \vee \chi^{q}_{j} \vee \cdots \vee \chi^{q}_{k}$, and so on. Due to the 
commutativity of the disjunction, this does not change $\hat{v}(\psi_{q})$. Now observe that it is always possible to find $x^{p}_{i}$ and $x^{q}_{j}$ 
such that $v(x^{p}_{i}) \neq v(x^{q}_{j})$. In fact, there are 16 possibilities of combining values $\hat{v}(\chi^{p}_{i})$ and $\hat{v}(\chi^{q}_{j})$
 because either $\chi^{p}_{i} = x^{p}_{i}$ and $\chi^{q}_{j} = x^{q}_{j}$, $\chi^{p}_{i} = \neg x^{p}_{i}$ and $\chi^{q}_{j} = x^{q}_{j}$, 
$\chi^{p}_{i} = x^{p}_{i}$ and $\chi^{q}_{j} = \neg x^{q}_{j}$ or $\chi^{p}_{i} = \neg x^{p}_{i}$ and $\chi^{q}_{j} = \neg x^{q}_{j}$. 
Since we take an $\chi^{p}_{i}$ such that $\hat{v}(\chi^{p}_{i}) = 1$, we just look at one of these possibilities that are compatible with the 
form of $\chi^{p}_{i}$ and for which $\hat{v}(\chi^{q}_{j}) = 1$ but  $v(x^{p}_{i}) \neq v(x^{q}_{j})$. For this reason, it can be supposed that
 $v(x^{p}_{i}) \neq v(x^{q}_{j})$. In this way, because $v(x^{p}_{i}) \neq v(x^{q}_{j})$, the product of the $a_{ii}$-element of 
$(((I_{k} - | v(x^{p}_{1}) \cdots v(x^{p}_{k}) \rangle \langle v(x^{p}_{1}) \cdots v(x^{p}_{k}) |) \otimes I_{n-k})$ and the $i$-element of 
$| v(x^{p}_{1}) \cdots v(x^{p}_{k}) \rangle$ is equal to zero. Indeed, this element $a_{ii}$ is just $v(x^{p}_{i}) \cdot v(x^{p}_{i})$ and the
 $i$-element of $| v(x^{p}_{1}) \cdots v(x^{p}_{k}) \rangle$ is $v(x^{p}_{i})$, and so $v(x^{p}_{i}) = 0$. Since $v(x^{p}_{i}) \neq v(x^{q}_{j})$,
 this means that $v(x^{q}_{j}) = 1$ and, consequently, $((I_{k} - | v(x^{q}_{1}) \cdots v(x^{q}_{k}) \rangle \langle v(x^{q}_{1}) 
\cdots v(x^{q}_{k}) |) \otimes I_{n-k}) | v(x^{q}_{1}) \cdots v(x^{q}_{k}) \rangle \neq 0$. With a similar argument we show that if 
$((I_{k} - | v(x^{p}_{1}) \cdots v(x^{p}_{k}) \rangle \langle v(x^{p}_{1}) \cdots v(x^{p}_{k}) |) \otimes I_{n-k}) | v(x^{p}_{1}) 
\cdots v(x^{p}_{k}) \rangle \neq 0$ then $((I_{k} - | v(x^{q}_{1}) \cdots v(x^{q}_{k}) \rangle \langle v(x^{q}_{1}) \cdots v(x^{q}_{k}) |) 
\otimes I_{n-k}) | v(x^{q}_{1}) \cdots v(x^{q}_{k}) \rangle = 0$.
\epr

Given what was said above, it can be derived from Proposition \ref{proposition} that $k$-$QSAT$ is not an adequate generalization of 
$k$-$SAT$ as far as the logical relationship between them is concerned.

\section{Conclusion}

In this paper, the logical relationship between $SAT$ and $QSAT$ was made explicit. It was shown that the connection between them 
is only superficial and not deep enough to allow a direct comparison between $NP$ and $QMA$. This result raises the question: Is there 
a $QMA$-complete problem that, from a logical point of view, is an appropriate quantum generalization of $SAT$?

The same limitations exhibited here pertaining SAT and QSAT also are applicable to the problems studied in \cite{liu2006} as well as in 
\cite{laumann2010} pertaining the relationship between PSAT and QSAT. Therefore, the existing quantum versions of the satisfiability problem do 
not allow an adequate logical analysis of the relationship between quantum and classical time-complexity classes.

This does not permit us, however, to affirm that all versions of $QSAT$ are inappropriate to compare $NP$ and $QMA$. Moreover, it is possible 
that $QSAT$ itself could be used for this aim. The point is that, although the existing quantum generalization of $SAT$ could seem to be 
analogous to it, they have in fact a logical formulation that is essentially different from $SAT$, the original problem.

\bibliographystyle{eptcs}

\bibliography{araujo-lsfa}

\end{document}